\documentclass[12pt]{spieman}
\usepackage{amsmath,amsfonts,amssymb}
\usepackage{graphicx}
\usepackage{setspace}
\usepackage{tocloft}
\usepackage{lineno}
\usepackage[colorlinks=true, allcolors=blue]{hyperref}
\usepackage{algorithmic}
\usepackage{algorithm, setspace}
\usepackage{multirow}
\usepackage{bm}
\usepackage{diagbox}

\usepackage{fancyhdr}
\usepackage{booktabs}

\def\PAnumber{2025-5579}
\fancypagestyle{empty}{
  \fancyhf{} 
  \fancyfoot[C]{\fontsize{11}{11}\selectfont Approved for public release; distribution is unlimited. Public Affairs release approval \# \PAnumber .}
  
}

\title{WindDensity-MBIR: Model-Based Iterative Reconstruction for Wind Tunnel 3D Density Estimation}

\author[a,*]{Karl J. Weisenburger}
\author[a]{Gregery T. Buzzard}
\author[b]{Charles A. Bouman}
\author[c]{Matthew R. Kemnetz}
\affil[a]{Purdue University, Department of Mathematics, West Lafayette, Indiana 47907, USA}
\affil[b]{Purdue University, Departments of Electrical and Computer Engineering and Biomedical Engineering, West Lafayette, Indiana 47907, USA}
\affil[c]{Air Force Institute of Technology, Department of Engineering Physics, Wright-Patterson AFB, OH 45433, USA}

\newcommand{\RR}{{\mathbb R}}
\newcommand{\bx}{\boldsymbol{x}}
\newcommand{\by}{\boldsymbol{y}}

\cftpagenumbersoff{figure}
\cftpagenumbersoff{table} 
\begin{document} 

\maketitle

\begin{abstract}
Experimentalists often use wind tunnels to study aerodynamic turbulence, but most wind tunnel imaging techniques are limited in their ability to take non-invasive 3D density measurements of turbulence. Wavefront tomography is a technique that uses multiple wavefront measurements from various viewing angles to non-invasively measure the 3D density field of a turbulent medium. 
Existing methods make strong assumptions, such as a spline basis representation, to address the ill-conditioned nature of this problem.   
We formulate this problem as a Bayesian, sparse-view tomographic reconstruction problem and develop a model-based iterative reconstruction algorithm for measuring the volumetric 3D density field inside a wind tunnel.  We call this method WindDensity-MBIR and apply it using simulated data to difficult reconstruction scenarios with sparse data, small projection field of view, and limited angular extent. WindDensity-MBIR can recover high-order features in these scenarios within 10\% to 25\% error even when the tip, tilt, and piston are removed from the wavefront measurements.

\end{abstract}


{\noindent \footnotesize\textbf{*}Karl J. Weisenburger, \linkable{kweisen@purdue.edu} }

\begin{spacing}{2}

\section{Introduction}
\label{sec:intro}

Wavefront sensing is a class of imaging techniques that can be used to detect variations in a 3D density field due to high-speed airflow\cite{AeroOpticReview}. When a coherent light source illuminates a turbulent airflow, changes in refractive index due to variations in density can affect the optical path length (OPL) of the light. Wavefront sensing techniques such as digital holography or Shack-Hartmann wavefront sensors can precisely measure the optical path difference (OPD) over a coherent beam's aperture, allowing researchers to non-invasively detect density variations for a turbulent airflow. However, OPD is a path-integrated measurement along a line of sight and so does not yield point estimates of the underlying 3D density. Alternatively, wind tunnel flow seeding can provide 3D point information, but it is invasive and difficult to perform \cite{FlowSeedingReview}. Computational fluid dynamics (CFD) is non-invasive and can model the 3D flow dynamics, but CFD simulations often fail to match experimental tests\cite{CFDKalenskyandJumper, CFDAndrewandThomas}.

Wavefront tomography is an alternative non-invasive method to estimate the 3D density field of a turbulent volume \cite{hesselink2001optical}. While similar to parallel-beam x-ray tomography, wavefront tomography for volumetric wind tunnel reconstruction is challenging is due to sparse views, limited field of view, limitations in angular extent, and incomplete projection information. To capture the dynamic flow field of wind tunnel turbulence, all wavefront measurements must be acquired at the same time. 
The physical constraints of imaging inside a wind tunnel limit the number and angular extent of wavefront measurements that can be obtained simultaneously, turning this into a limited-angle, sparse view tomography problem. Furthermore, 2D OPD measurements are not complete projections of the turbulence refractive index because they lack the low order tip, tilt, and piston (TTP) information. By definition, OPD is blind to the mean OPL over the aperture (i.e., piston), and mechanical disturbances often contribute an unknown tip-tilt to the wavefront, which effectively obscures the tip-tilt contribution from the turbulence. In some cases, it is possible to recover the complete projection information if the projection field of view (FOV) is wider than the medium of turbulence\cite{Zhang:93}. However, for realistic wind tunnel experiments this is not possible because the FOV is limited by the beam diameter. Lastly, most wind tunnels have physical viewing constraints that also restrict the angular extent of the projections, making it especially difficult to resolve features along one axis of the reconstruction. 

Wavefront tomography has been explored extensively in the field of microscopy for the goal of reconstructing the 3D sample refractive index \cite{DH-tomo}. While many of the methods used in this field are designed for limited angle or sparse viewing scenarios, they do not consider scenarios where the object of reconstruction is significantly wider than the projection FOV. Furthermore, since the FOV is wider than the sample, the additional projection information outside the sample (i.e., through the background reference medium) enables the recovery of the complete projection information (i.e., TTP included) for the sample. 

Several wavefront tomography methods have been developed for wind-tunnel turbulence reconstruction, but none of them use a Bayesian model-based reconstruction framework, which is well established in the field of x-ray tomography for regularization and artifact removal\cite{MBIR_review}. Some reconstruction methods implicitly perform regularization by parameterizing the reconstruction with a low dimensional smooth orthogonal basis\cite{Zhang:93,ChaoTian}, but these techniques do not use an explicit Bayesian prior model. Some sophisticated approaches have been developed to account for missing projection information, which either incorporate additional reference information\cite{Zhang:93,ChaoTian} or an iterative update loop \cite{Vest,SoyoungS.Cha}, but in all cases they do not use an explicit Bayesian prior model to smooth out the artifacts resulting from the unknown or incorrect projection information.

Furthermore, most of the methods designed for turbulence reconstruction are not designed for difficult imaging scenarios where the issues of sparse view, limited angle, small FOV, and unknown projection TTP are all present. For instance, some methods assume that the turbulent medium is radially symmetric\cite{Vest,ChaoTian}, which means that a single measurement can reveal all 180 degrees of projection information. Alternatively, other methods assume that the turbulence stream induces a steady flow field\cite{Soller}, which means that many measurements can be taken over the course of an extended time interval. While some investigations into wind tunnel wavefront tomography did assume scenarios with a severely limited number of measurements, none of them considered cases where the angular extent was also severely limited (i.e., less than 90 degrees)\cite{RobertLJohnson,Soller,Vest,SoyoungS.Cha,Zhang:93,ChaoTian,McMackin:97}. Furthermore, most experiments and simulations of wavefront tomography assume that each projection FOV is significantly wider than the medium of turbulence\cite{RobertLJohnson,Soller,Vest,SoyoungS.Cha,ChaoTian,McMackin:97}. Y. Zhang and G. A. Ruff specifically investigated the issue of small FOV relative to the turbulent medium and developed a method for estimating the unknown TTP\cite{Zhang:93}. However, their technique assumes additional reference information, such as temperature probe measurements throughout the turbulent medium, which is invasive and difficult to obtain in practice. 

Some wavefront tomography techniques overcome the issues of sparse measurement data, small FOV, and limited angular extent by either using layered models of turbulence\cite{multiconjugateadaptiveopticsastronomy,DeepTomography,SingleGuideStarTomography,SamThurman} or lower-dimensional basis representations \cite{SamThurman,Zhang:93,ChaoTian} to reduce the complexity of the reconstruction problem. Tomographic algorithms used in multi-conjugate adaptive optics systems model atmospheric turbulence as several layered planes distributed at various elevations\cite{multiconjugateadaptiveopticsastronomy,DeepTomography,SingleGuideStarTomography}. Similarly, Klee et al.\cite{SamThurman} demonstrated the experimental feasibility of tomographic wavefront sensing for reconstructing two phase screens aligned along an optical bench. Their reconstruction algorithm also assumed a spline basis representation for each phase screen and solved for the basis coefficients. While layered models of turbulence are appropriate for many applications they are not designed for volumetric reconstruction or capturing the spatially correlated features of wind tunnel turbulence. Furthermore, methods that use lower-dimensional basis representations are better posed, but their resolving capacity is limited to the features in the basis span.

In this paper, we introduce WindDensity-MBIR \cite{Repo}, a model-based iterative reconstruction (MBIR) method for non-invasive estimation of the volumetric density field inside a wind tunnel. We formulate the volumetric wavefront tomography problem as parallel beam tomography and use existing parallel beam MBIR tomography algorithms and software\cite{mbirjax-2024} to estimate the unknown densities. We show that WindDensity-MBIR can recover high order features within 10\% to 25\% error even when there are few measurements, the angular extent is severely limited, and the measurements have TTP removed. We compare WindDensity-MBIR to scale-corrected filtered back projection (FBP) and find that MBIR significantly outperforms FBP. We also investigate the effect of optical configuration on reconstruction quality and determine that reconstruction accuracy improves only when the maximum angular extent and the number of views increase simultaneously. Lastly, we explore the deleterious effects of using non-ideal tip-tilt removed OPD measurements by performing reconstructions with ideal OPL measurements and comparing the two cases. In our simulations, roughly 95\% of the additional error due to the unknown TTP is contained in lower order Zernike modes (of radial degree 2 or less), which implies that WindDensity-MBIR applied for wavefront tomography is able to recover high order modes even using TTP-removed OPD measurements.

\section{Methodologies} 
\subsection{Physical Model}
\label{sec: physical model}

The physical quantity of interest is the 3D density field $\boldsymbol{\rho}$ inside the wind tunnel. However, wavefront measurements are projections of the refractive index field $\boldsymbol{n}$, which is related to the density field $\boldsymbol{\rho}$ via the Gladstone-Dale equation\cite{GladeStoneOG,GladeStoneExperimental}: 
\begin{equation}
\label{eq:Refractiveindex2density}
\boldsymbol{\rho}(\vec{r})=\frac{\boldsymbol{n}(\vec{r})-1}{K_{GD}},
\end{equation} 
where $K_{GD}$ is the Gladestone-Dale constant and $\vec{r}$ denotes a point in the volume. Thus, because wavefront measurements are projections of refractive index, which is directly related to density via Eq.~\eqref{eq:Refractiveindex2density}, for the rest of this paper we will focus on reconstructing the refractive index field $\boldsymbol{n}$. 

\begin{figure}[t]
    \centering
    \includegraphics[width=0.9\textwidth]{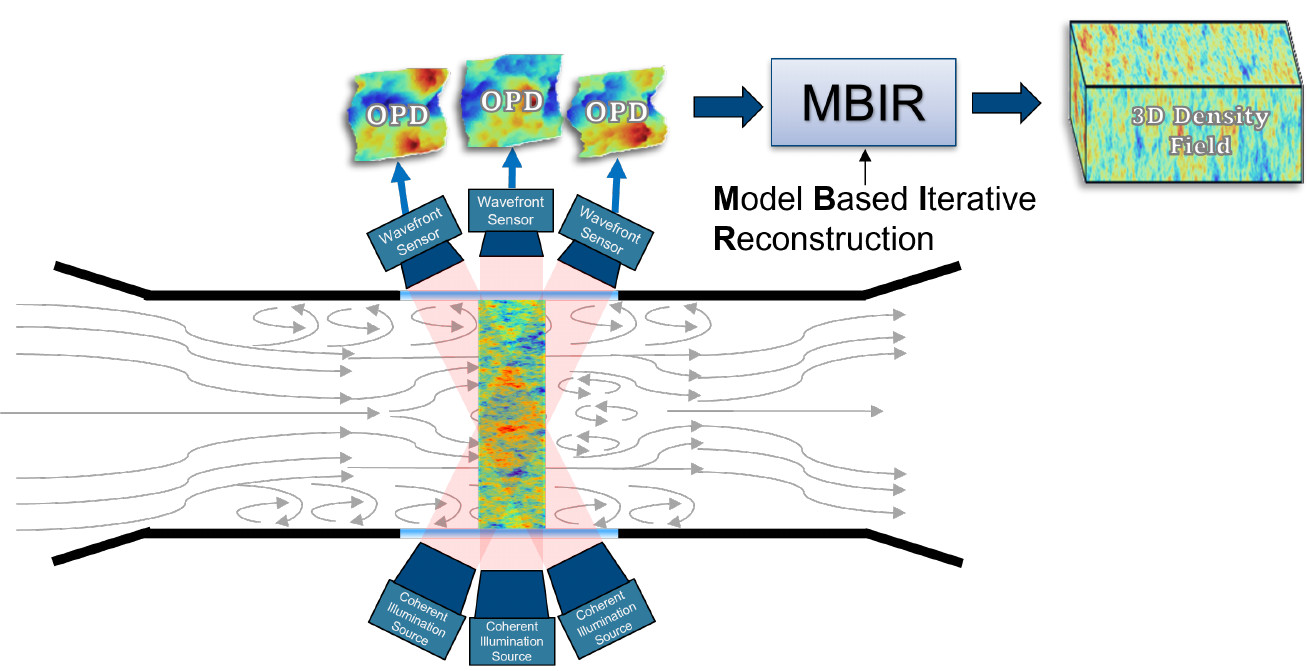}
    \caption{Notional depiction of a wind tunnel wavefront tomography set up.}\label{Fig: wind tunnel tomography}
\end{figure}

A basic wavefront tomography measurement system would entail several individually collimated coherent beam sources along with several wavefront detection systems. A wavefront detection system would include either a set of Shack–Hartmann wavefront sensors or perhaps several digital holography set-ups each with their own reference beam. Figure~\ref{Fig: wind tunnel tomography} provides a notional representation of a possible wavefront tomography set-up.
 
We denote a single projected ray through the volume as $\gamma_j$, where $j$ indexes a particular pixel in a particular view.
Then we model the OPL for a single ray as equal to the integral of the refractive index $\boldsymbol{n}$ along the path $\gamma_j$ given by
\begin{equation}
\label{eq:OPL}
OPL_j = \int_{\gamma_j} \boldsymbol{n}(\vec{r}) d\vec{r} \ ,
\end{equation} 
where $\boldsymbol{n}(\vec{r})$ denotes the index of refraction at location $\vec{r}$.

To discretize Eq.~\eqref{eq:OPL}, we replace the continuous refractive index volume $\boldsymbol n$ with a discrete vectorized volume of $N$ pixels $\boldsymbol{x} \in \RR^{N}$. We replace the continuous integral along the path $\gamma_j$ with a linear forward projection given by the inner product with the column vector $A_j\in\RR^{N}$. Thus the discrete $OPL$ is given by
\begin{equation}
\label{eq:Discrete Approximation}
A_j^t\boldsymbol{x} \approx \int_{\gamma_j} \boldsymbol{n}(\vec{r})\cdot d\vec{r}=OPL_j.
\end{equation} 

If the projected beams are collimated, the turbulence is weak, and the overall projection distance is short, then we can approximate the behavior of each projected beam as a bundle of parallel light rays that each follow a straight line path through the volume of turbulence. Under this approximation, the linear operator $A_j^t$ is the parallel beam forward projection for a single measured pixel in a single view. Thus, the complete forward model for the entire system of views is given by
\begin{equation}
\label{eq:Full Physical model}
\boldsymbol{y}= \boldsymbol{Ax}+\boldsymbol{W}
\end{equation} 
where $\boldsymbol{y}$ is the set of all OPL measurements, $\boldsymbol{A}=[A_0,...,A_{M-1}]^t\in\RR^{M\times N}$ is the forward projection operator for all views, and $\boldsymbol{W}\sim N(0,\sigma_yI)$ is additive white Gaussian noise. Here $M$ is the total number of pixels across all beam apertures. Also, note that $\boldsymbol{y}$ contains all the measurements for all projection views and all angles.

\subsection{Reconstruction Algorithm}

For a typical model-based approach, the final reconstruction $\boldsymbol{\hat x}$ is the maximum a posteriori (MAP) estimate of the ground truth $\boldsymbol{x}$. The MAP estimate is the reconstruction that jointly minimizes the forward model and prior model:
\begin{equation}
\label{eq:MAP_estimate}
\hat{\boldsymbol{x}}= \arg\min_{\boldsymbol{x}} \{ -\log p(\boldsymbol{y}|\boldsymbol x) -\log p(\boldsymbol{x}) \} = \arg\min_{\boldsymbol{x}} \{ f(\boldsymbol{x};\boldsymbol y) + h(\boldsymbol{x}) \},
\end{equation}
where $f(\bx; \by)$ promotes the fit of $\boldsymbol{A} \bx$ to the measured sinogram $\by$, and $h(\bx)$ represents the prior information on $\bx$.

Using the parallel-beam framework established in Sec. \ref{sec: physical model}, which assumes Gaussian measurement noise, we can write the forward model $f(\boldsymbol{x};\boldsymbol y)$ for reconstructing the index of refraction $\boldsymbol{x}$ as
\begin{equation}
    \label{eq:Forward Model}
    f(\boldsymbol{x};\boldsymbol y)= -\log p(\boldsymbol y |\boldsymbol x) =\frac{1}{2\sigma_{y}^{2}} ||\boldsymbol y -\boldsymbol{Ax}||^2_\Lambda + C,
\end{equation}
where $C$ is a constant independent of $\boldsymbol x$, $\sigma_{y}$ is the assumed standard deviation of the measurement noise $\boldsymbol W$, and $\Lambda \in\RR^{M\times M}$ is a diagonal weight matrix (dependent on the optical configuration and $\by$) that defines the uncertainty for each OPL projection $y_j$.

For our prior model $h(\boldsymbol{x})$ we chose the generalized Gaussian Markov random field (GGMRF)\cite{https://doi.org/10.1118/1.2789499}. Technically, to achieve the true MAP estimate, the prior model must equal (up to a constant) the negative log prior distribution for the volume being reconstructed $h(\boldsymbol x)=-\log p(\boldsymbol{x})+C'$. However, it is often sufficient if $h(\boldsymbol{x})$ simply captures some reasonable statistical assumptions about the volume being reconstructed. In our case, we assume that the volume $\boldsymbol{x}$ is smooth, having minimal variation between most adjacent voxels. The GGMRF model can implement these assumptions while also providing flexibility to tune the amount regularization. We use the existing MBIRJAX package\cite{mbirjax-2024} to solve Eq.~\eqref{eq:MAP_estimate} since it has implementations of the parallel beam forward model in Eq.~\eqref{eq:Forward Model} and the GGMRF prior model. The MBIRJAX package solves Eq.~\eqref{eq:MAP_estimate} with vectorized coordinate descent\cite{VCD}.
 
 \subsection{Using Non-Ideal Tip-tilt Removed Optical Path Difference}
 While the reconstruction problem given by Eq.~\eqref{eq:MAP_estimate} assumes that the measurement for each beam is OPL, this is not experimentally feasible. Standard wavefront detection techniques, such as digital holography or Shack–Hartmann wavefront sensors, measure only the wavefront OPD, i.e., the mean-removed OPL. Furthermore, for measurements through wind tunnel turbulence, most OPD sensing techniques cannot recover the tip-tilt (i.e., the linear trend across the aperture) from the turbulence because this is obscured by the tip-tilt contributions from mechanical disturbances. This means that for experimental measurements of turbulence in wind tunnels, the tip-tilt is usually removed from the OPD. The tip-tilt removed OPD measurement, $\text{OPD}_{\text{TT}}$, can be expressed as 
 \begin{equation}
\label{eq: TTP Removal}
OPD_{TT}(u,v)=OPL(u,v)-(au+bv+c)
\end{equation}
where
\begin{equation}
\label{eq: TTP fitting}
(a,b,c)=\arg\min_{a,b,c}{ \| OPL(u,v)-(au+bv+c)\|^2}.
\end{equation}

Thus, for the primary analysis of our algorithm, we used simulated $\text{OPD}_{\text{TT}}$ measurements as the input for our algorithm instead of OPL. Because the forward model given by Eq.~\eqref{eq:Forward Model} assumes that the measurements $\boldsymbol y$ are OPL, using $\text{OPD}_{\text{TT}}$ instead of OPL introduces model mismatch and degrades reconstruction quality. To determine the effects of model mismatch due to using $\text{OPD}_{\text{TT}}$ instead of OPL measurements, we also performed a secondary analysis in Sec. \ref{sec: OPL Results} where we simulated reconstructions with OPL measurements. Comparing the error between reconstructing with $\text{OPD}_{\text{TT}}$ and reconstructing with $\text{OPL}$ indicated that the additional error due model mismatch is mostly contained in lower order Zernike modes.

\subsection{Simulation and Validation Procedure}
\subsubsection{Atmospheric phase volume generation}

To validate the effectiveness of WindDensity-MBIR, we randomly generated 3D phase volumes based on the Kolmogorov phase power spectral density (PSD). We chose to use Kolmogorov atmospheric turbulence for our simulations because it is well understood and straightforward to model. 

To generate 3D phase volumes, we extend the commonly used Fourier transform method for generating 2D atmospheric phase screens\cite{alma9972291887508496} to generate 3D volumes. We make use of the modified von K\'arm\'an phase PSD:
\begin{equation}
\label{eq:Phase PSD}
\Phi^{mvK}_\phi(\kappa) = 0.49r_0^{-\frac{5}{3}} \frac{\exp\{-\kappa^2/\kappa^2_m\}}{(\kappa^2+\kappa_0^2)^\frac{11}{6}}
\end{equation}
where $r_0$ is the Fried's coherence length in meters, $\kappa_m$ and $\kappa_0$ are the high and low frequency scale parameters, and $\kappa$ is the angular wave number, all in $m^{-1}$.

To generate a volume with physical dimensions $(L, W, H)$ in meters on a finite grid with PSD given by \eqref{eq:Phase PSD}, we first consider the phase volume $\boldsymbol{x}$ as a Fourier series:
\begin{equation}
\label{eq: Fourier Representation}
\boldsymbol{x}(\vec{r})=\sum_{\vec{n} \in \mathbb{Z}^3} c_{\vec{n}} \exp\left\{j2\pi\langle F \vec{n}, \vec{r}\rangle \right\}
\end{equation}
where $F=diag([1/L,1/W,1/H])$ is a diagonal matrix of the frequency spacings. 

 From here we model the Fourier coefficients as having Gaussian real and imaginary parts,
\begin{equation}
\label{eq: Fourier Coefficient}
\Re \{c_{\vec{n}}\} \sim N(0,\sigma_{\vec{n}}^2) \quad \textrm{and} \quad
\Im \{c_{\vec{n}}\} \sim N(0,\sigma_{\vec{n}}^2),
\end{equation}
each with variance $\sigma_{\vec{n}}^2$ specified by $\Phi^{mvK}_\phi$,

\begin{equation}
\label{eq: Variance}
 \sigma_{\vec{n}}^2= \frac{\Phi^{mvK}_\phi(\|F\vec{n}\|_2)}{LWH}.
\end{equation}
Following the approach used by Schmidt\cite{alma9972291887508496}, we implement Eqs. \eqref{eq: Fourier Representation} - \eqref{eq: Variance} in software with pseudo-random number generation and the fast Fourier transform. 

Importantly, we are not attempting to simulate statistically accurate ground truth refractive index volumes. The Kolmogorov phase PSD models the 2D phase fluctuations for a coherent beam passing through a medium of atmospheric turbulence. It does not model a volume of atmospheric refractive index. For the purpose of performing a basic test of WindDensity-MBIR, which was ultimately designed for aero-optical turbulence in a wind tunnel, our goal was simply to generate statistically independent volumes containing random isotropic turbulent structures that are reasonably sized. The modified von K\'arm\'an PSD for phase screens is appropriate for this goal because the inner and outer scale parameters ($L_0=2\pi/\kappa_0$ and $l_0=2\pi/\kappa_m$) allow us to easily control the size range of the turbulent structures. For all tests done in this paper, we used $r_0=0.5$ cm, $L_0= 2$ cm, and $l_0=0$.

\subsubsection{Simulating OPL and tip-tilt removed OPD measurements}

\begin{figure}[t]
\centering
    \includegraphics[width=0.96\textwidth]{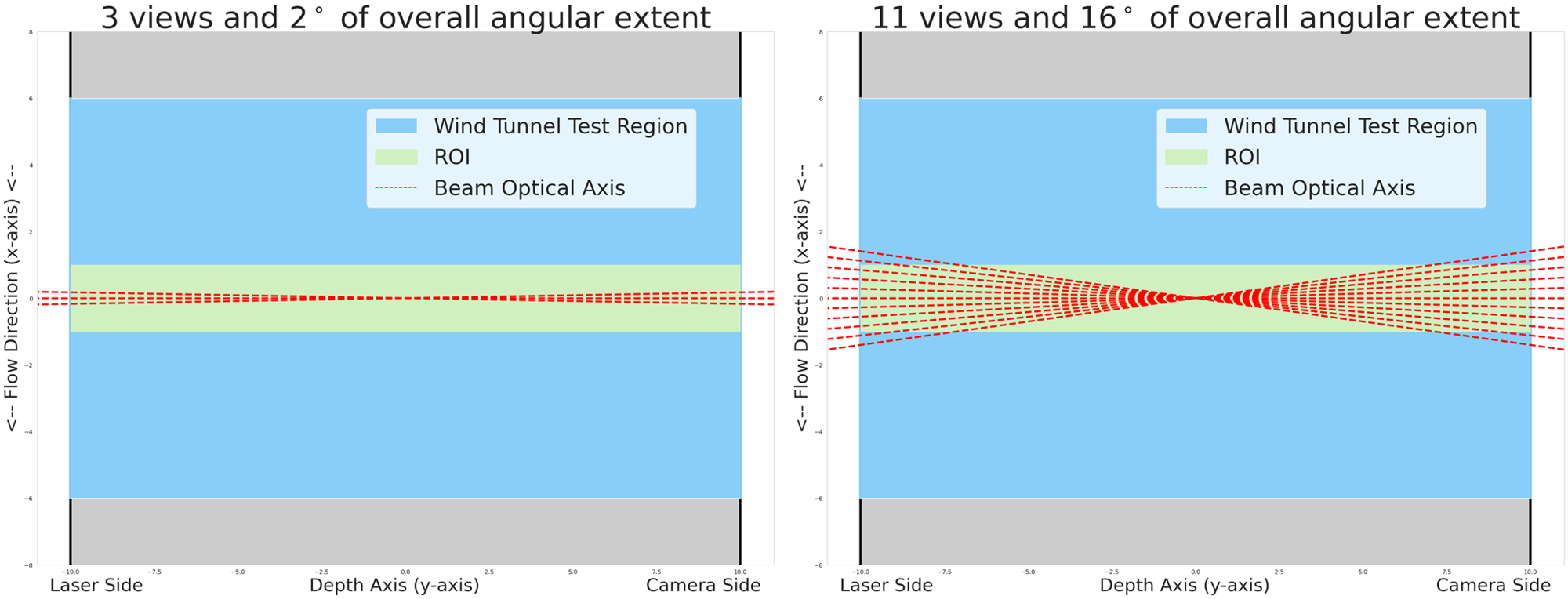}
    \newline
    (a) \hspace{7.5cm} (b)
    \caption{Overhead view of (a) the geometry with the fewest views and narrowest angular extent and (b) the geometry with the most views and widest angular extent.}\label{Fig: Geometries}
\end{figure}

For the results shown in this paper, we tested $40$ different optical configurations, varying both the number of beams and the overall angular extent. For the total number of beams, we tested $3$, $5$, $7$, $9$, and $11$ beams. For the total angular extent of we tested $2^\circ$, $4^\circ$, $6^\circ$, $8^\circ$, $10^\circ$, $12^\circ$, $14^\circ$, and $16^\circ$. In every case, the optical axes of the beams lie on a single plane perpendicular to the axis of rotation. Furthermore, for every configuration, the beam angles are uniformly distributed across the total angular extent, which means the angular separation between adjacent beams is always $\textit{angular extent}/ (\textit{number of views} - 1)$. Figure~\ref{Fig: Geometries} shows the beam paths for one geometry with the narrowest angular extent and fewest views and another geometry with the widest angular extent and most views.

For a given simulated volume of refractive index $\boldsymbol x$ and an optical configuration, computing the OPL and the tip-tilt removed OPD is relatively straightforward. Following Eq.~\eqref{eq:Discrete Approximation}, we compute the OPL forward projection for each beam as the integral of $\boldsymbol x$ along all rays parallel to the beam's optical axis. We use the MBIRJAX forward projection operator\cite{mbirjax-2024} to perform this computation. Then we set all values outside the beam's predefined FOV to zero.  For all results shown in this paper, we set the beam FOV to be a $2$ cm disk. Finally, we compute the $\textrm{OPD}_\textrm{TT}$ (tip-tilt removed OPD) by fitting a plane to the OPL and then subtracting it from the OPL (see Eq.~\eqref{eq: TTP fitting} and \eqref{eq: TTP Removal}).

\subsubsection{Resolution reduction and conventions for error analysis}

\begin{figure}[ht]
    \centering
    \includegraphics[width=0.7\textwidth]{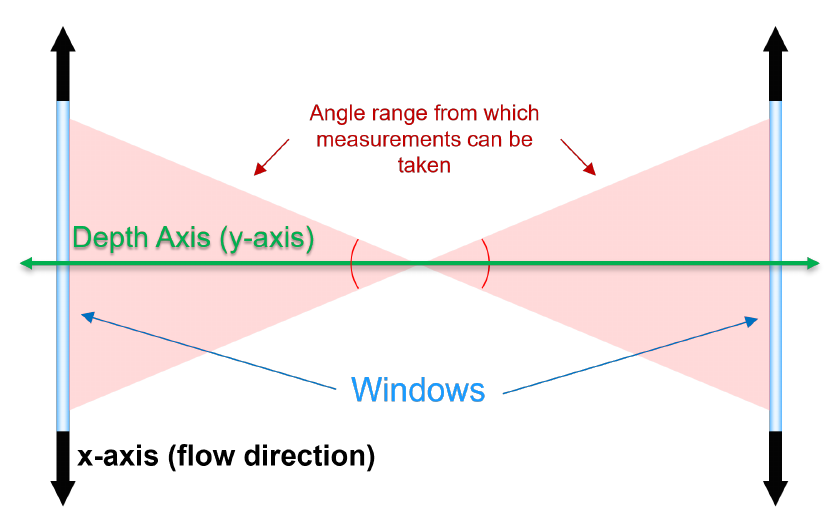}
    \caption{Schematic showing the depth axis relative to the location of the windows. The red shaded region designates range of possible view angles.}\label{Fig: Depth axis}
\end{figure}

Throughout this paper we use the term ``depth axis" to refer to the line that bisects the largest angular range from which we take our measurements. This axis is the same for all our tests and is perpendicular to both windows on either side of the wind tunnel test region (see Fig.~\ref{Fig: Depth axis}).

\begin{figure}[t]
    \centering
    \includegraphics[width=0.8\textwidth]{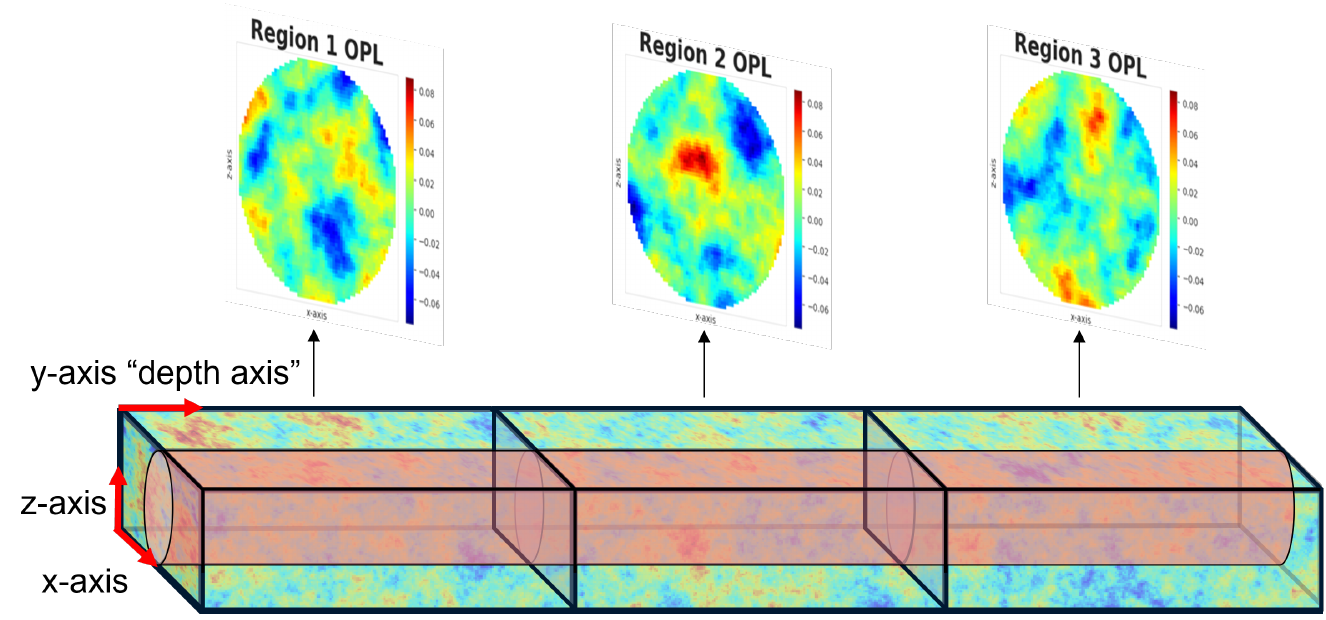}
    \caption{Example depiction of the integration process for reducing the resolution along the depth axis to 3 OPL planes. The windows in Fig.~\ref{Fig: Depth axis} correspond to the left and right ends of this volume.  The red cylinder represents our chosen reconstruction ROI, i.e., the path of the central beam.}\label{Fig: Block Averaging}
\end{figure}

For this paper, we restrict our region of interest (ROI) to the region of turbulence that the central beam passes through. For our set up, this region is a cylinder extending along the depth axis, perpendicular to both windows (see Fig.~\ref{Fig: Block Averaging}). All error computations and visualizations in this paper will be restricted to this region.

For visualization purposes (and to test the limits of depth axis resolution using WindDensity-MBIR) we often reduce the resolution along the depth axis by computing the OPL for a few regions along this axis. To do this we simply divide each full resolution volume (reconstruction and ground truth) into a specified number of equally sized regions along the depth axis and then integrate each of them along the depth axis. Figure~\ref{Fig: Block Averaging} shows a visualization of this process for the case of reducing the resolution to 3 OPL planes along the depth axis. Note that in some cases we will also consider the tip-tilt removed OPD contributions from each region, which we will refer to as $\text{OPD}_{\text{TT}}$ planes.

To evaluate the reconstruction accuracy relative to the simulated ground truth, we use stable range normalization for our normalized root mean squared error (NRMSE) computations. This means that we normalize the root mean squared error by dividing by the inner 90th interpercentile range of the ground truth volume.

\subsubsection{Zernike expansion}
\label{sec: zernike_expansion} 
\begin{figure}[t]
    \centering
    \hspace*{1.5cm}
    \includegraphics[width=0.5\textwidth]{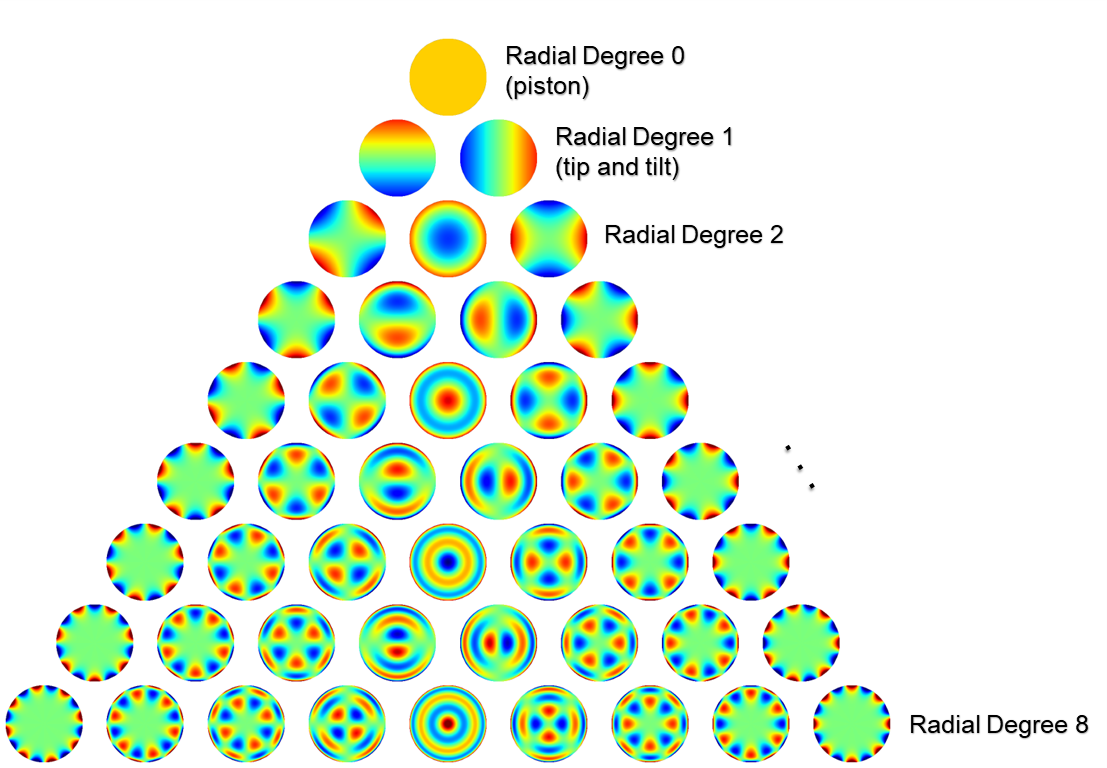}
    \caption{The first 45 Zernike modes ordered by radial degree.}\label{Fig: Zernike Degrees}
\end{figure}

Figure~\ref{Fig: Zernike Degrees} displays the first 45 Zernike modes, organized into 9 radial degrees, from the Zernike basis, which is an orthogonal basis using the $L_2$ inner product over the unit disk.  We use this basis in Sec.~\ref{sec: zernike1} and \ref{sec: zernike2} to analyze the low degree energy distribution of WindDensity-MBIR's reconstruction error. To do this in the case of known ground truth $\bx$ and reconstruction $\hat{\bx}$, we apply depth axis resolution reduction to $\bx$ and $\hat{\bx} - \bx$ as in the previous section.  We then project each resulting 2D region onto the span of one radial degree to obtain an error image, $\hat{\bx}_{d,j}^{err}$ for degree $d$ and region $j$.  We then report $\sum_j \|\hat{\bx}_{d,j}^{err}\|^2 / \sum_j \|\bx_j\|^2$ as the normalized mean-squared error for that degree.  

\section{Results}
\label{sec: Results}
In this section we investigate the performance of WindDensity-MBIR using a variety of evaluation methods.  For all of the simulations except the investigation of geometry configurations, we use the 7-view geometry with 8 degrees of angular extent.  This geometry is not numerically optimal, but we chose it as an approximate limit of physical feasibility. For convenience, we will refer to the geometry with $N$ views and $\theta$ angular extent as the ``$N$v,$\theta^\circ$-geometry".

\subsection{Comparison to Scale-Corrected FBP}

Scale-corrected FBP is a version of standard FBP that is scaled to better match the measured sinogram. The scale-corrected FBP is given by
 \begin{equation}
\label{eq: scalecorrfbp}
 \boldsymbol{\hat{x}}_{FBPcorrected}=\alpha\cdot\boldsymbol{\hat{x}}_{FBP}.
\end{equation}
where $\boldsymbol{\hat{x}}_{FBP}$ is the standard FBP reconstruction and the scaling coefficient $\alpha$ is given by
\begin{equation}
\label{eq: scaleCoefficient}
 \alpha= \frac{\langle\boldsymbol{y},\boldsymbol A \boldsymbol{\hat{x}}_{FBP}\rangle}{\|\boldsymbol A \boldsymbol{\hat{x}}_{FBP}\|_2^2}.
\end{equation}
Thus, computing the scale-corrected FBP reconstruction requires an additional forward projection of the standard FBP reconstruction (i.e., $\boldsymbol A \boldsymbol{\hat{x}}_{FBP}$) but in return gives the scaling that minimizes the mean-squared error between the sinogram and the projected reconstruction.

\begin{figure}[t]
    \centering
    \includegraphics[width=0.6\textwidth]{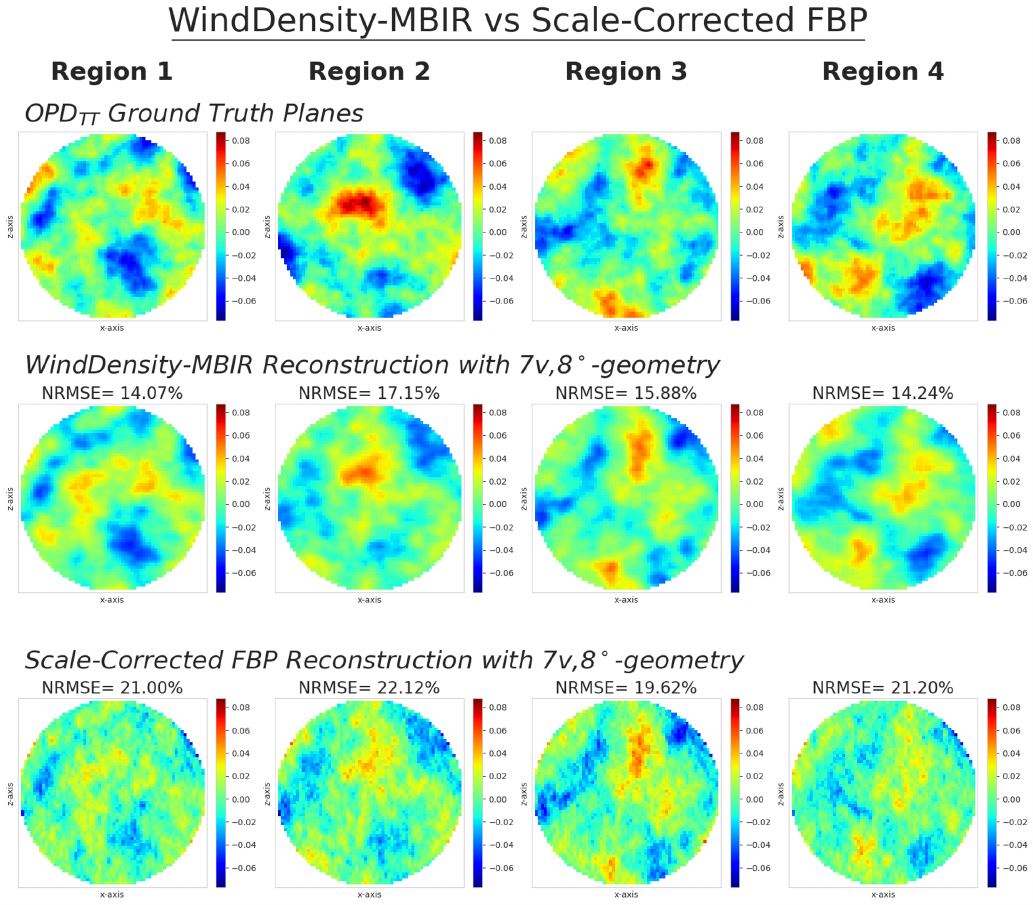}
    \caption{Effect of reconstruction method on reconstruction quality.  Example reconstructions of 4 $\text{OPD}_{\text{TT}}$ planes along the depth axis using the 7v,8$^\circ$-geometry. The top row shows ground truth, the middle row shows WindDensity-MBIR, and the bottom row shows scale-corrected FBP.  The limited measurements lead to high-frequency artifacts in the scale-corrected FBP reconstruction.  }\label{Fig: FBP Comparison}
\end{figure}

Figure \ref{Fig: FBP Comparison} shows example reconstructions comparing scale-corrected FBP to WindDensity-MBIR using the 7v,8$^\circ$-geometry. Even with scale correction, FBP produces poor results in comparison to MBIR because it is ill-suited for sparse reconstruction.

Table \ref{tab: FBP Table} shows the average NRMSE of 3 methods for reconstructing 4 $\text{OPD}_{\text{TT}}$ planes, using 3 representative geometries. The average NRMSE was computed for 100 different reconstructions. Here we see that scale-corrected FBP has lower NRMSE than standard FBP, by a factor of about 10, but is still inferior to MBIR.

\begin{table}[t]
    \centering
    \caption{NRMSE of FBP and MBIR for Reconstructing 4 $\text{OPD}_{\text{TT}}$ planes, averaged over 100 samples}
    \label{tab: FBP Table}
    \begin{tabular}{|c||c|c|c|c|}
        \hline
        \diagbox{Method}{Geometry} & 3v,2$^\circ$  & 7v,8$^\circ$ & 11v,16$^\circ$\\
        \hline 
        FBP & 5.689 & 2.191 & 1.734 \\
        Scale-Corrected FBP & 0.266 & 0.227 & 0.219 \\
        WindDensity-MBIR (ours) & 0.215 & 0.179 & 0.168 \\
        \hline
    \end{tabular}
\end{table}

\subsection{Reconstruction Error Analysis}
\label{sec: OPD Results}
\subsubsection{Error as a function of viewing geometry}

\begin{figure}[t]
\centering
    \includegraphics[width=0.98\textwidth]{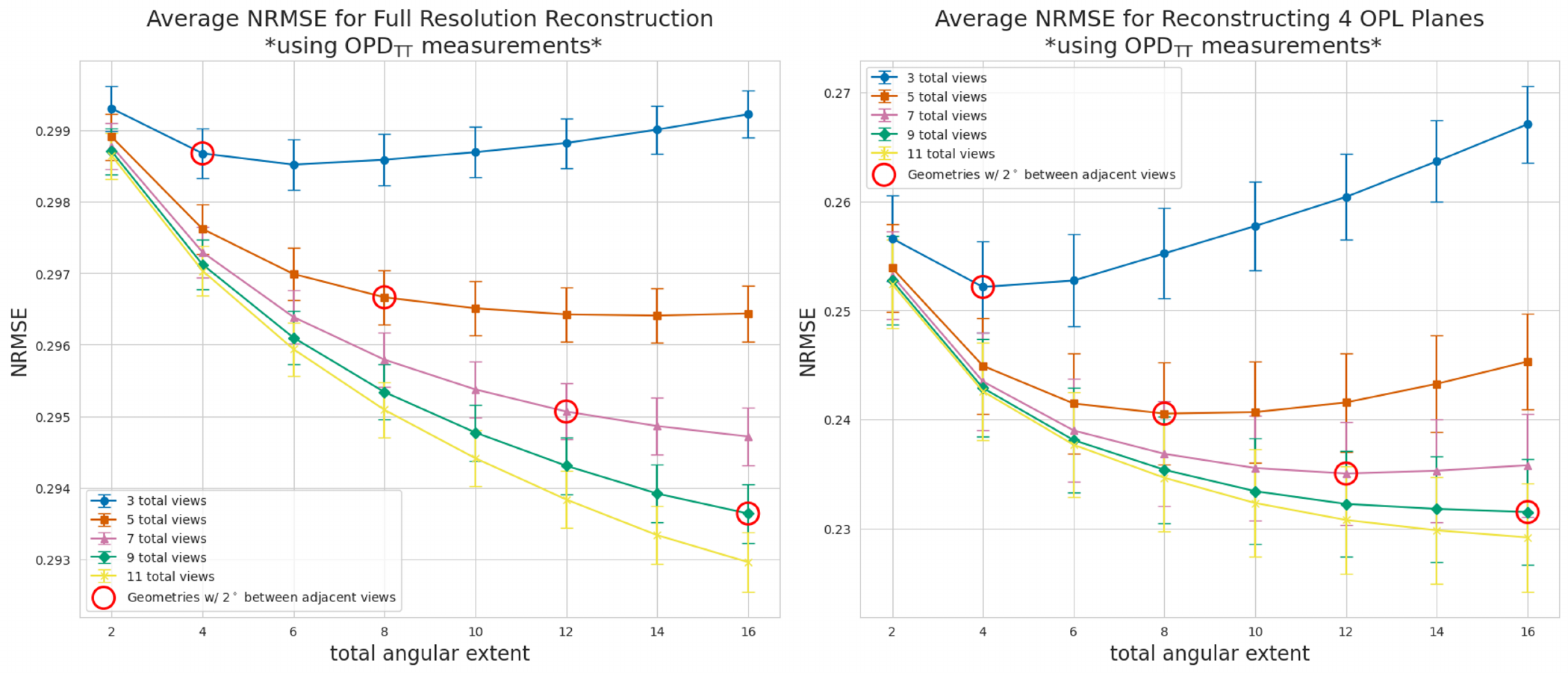}
    \newline
    \hspace*{0.45cm}(a) \hspace{7.3cm} (b)
    \caption{Plots of average NRMSE relative to the total angular extent for the viewing configuration: (a) reconstructing full resolution and (b) reconstructing 4 OPL planes. The red circles designate geometries with 2 degrees of angular separation between adjacent views and highlight the decrease in error as angular extent increases simultaneously with number of views. }\label{Fig: NRMSErelativetoOpticalConfig}
\end{figure}

Figure~\ref{Fig: NRMSErelativetoOpticalConfig} shows the NRMSE of WindDensity-MBIR for each of the 40 viewing configurations, averaged over reconstructions of 100 simulated volumes of turbulence. Plot (a) shows the NRMSE for a full resolution reconstruction (voxel depth equal to voxel width) and thus reveals the essential relationship between the viewing configuration and reconstruction accuracy. Plot (b) shows the NRMSE for a resolution of 4 OPL planes, which is a more feasible resolution for achieving high accuracy. While the $y$-axis scaling is different between the (a) and (b) plots, the overall trend is the same. Both plots highlight the importance of balancing angular extent with view overlap. Particularly in (b), the reconstruction quality consistently improves only when we increase the number of views and the total angular extent simultaneously. 

Figure~\ref{Fig: NRMSErelativetoOpticalConfig} indicates that for $2^\circ$ and $4^\circ$ of overall angular extent, increasing the views can only marginally improve the accuracy. Furthermore, when the number of views is limited to 3 or 5, increasing angular extent can actually degrade the reconstruction accuracy. This is because increasing the overall angular extent without increasing the number of views is equivalent to increasing the total number of unknowns without increasing the number of measurements. 

\begin{figure}[t]
\centering
    \includegraphics[width=0.98\textwidth]{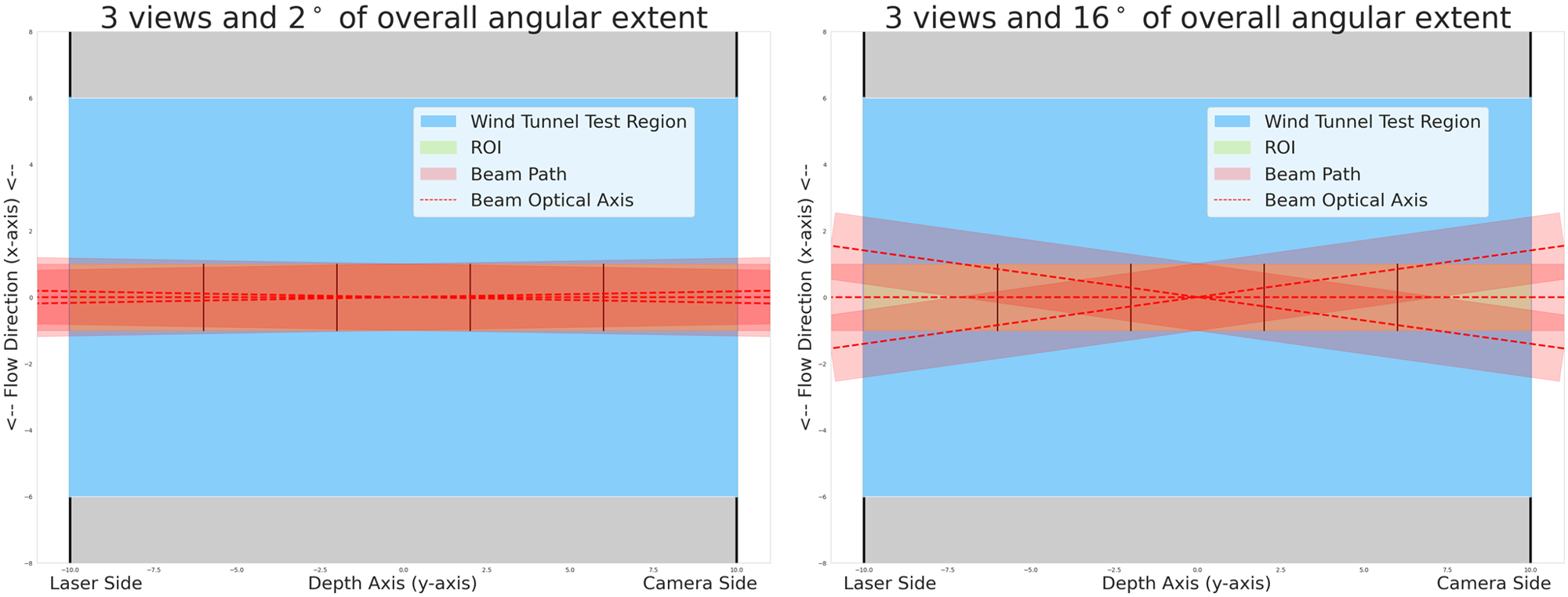}
    \newline
    (a) \hspace{7.5cm} (b)
    \caption{Overhead view of two edge case geometries: (a) the 3v,2$^\circ$-geometry and (b) the 3v,16$^\circ$-geometry (right). Note, the ROI in green is broken into the 5 regions that are reduced to 5 OPL planes for the experiment.}\label{Fig: NarrowandWideGeo}
\end{figure}

Figure~\ref{Fig: NarrowandWideGeo} shows two edge case geometries (3v,2$^\circ$ and 3v,16$^\circ$) that illustrate the fundamental trade-off between angular extent and view overlap. To investigate this trade-off, we reconstructed 3000 different volumes of turbulence for each of these viewing configurations. We then reduced the resolution to 5 OPL planes along the depth axis and computed the error individually for each plane. 

\begin{figure}[t]
\centering
    \includegraphics[width=0.9\textwidth]{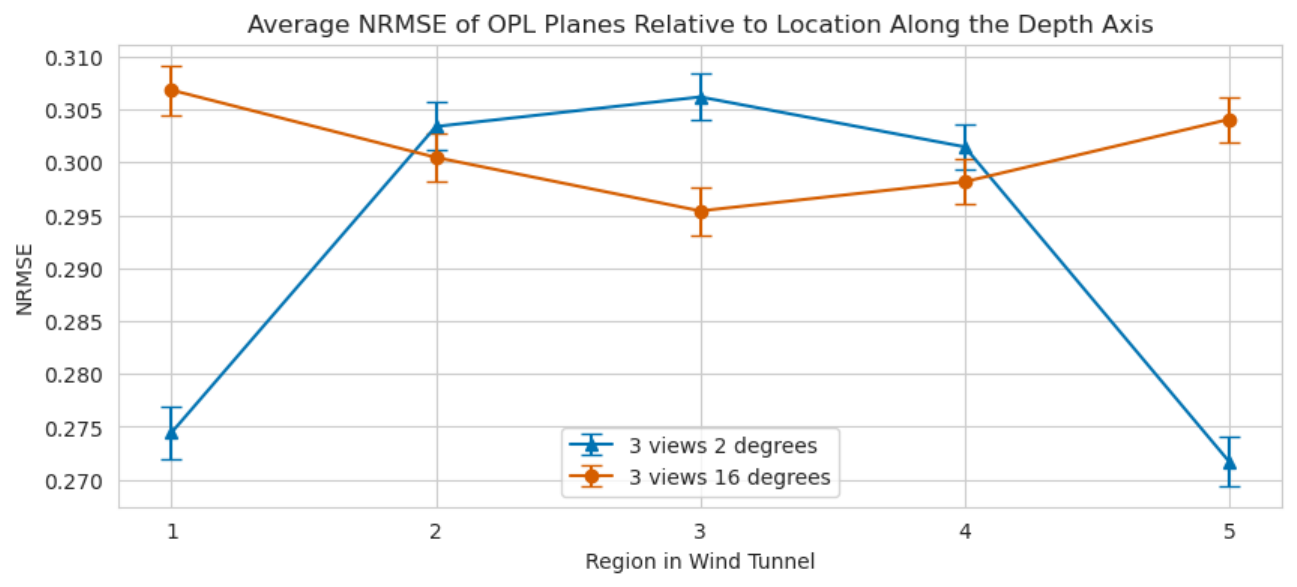}
    \caption{NRMSE as a function of region along the depth axis, averaged over 3000 reconstructions. Error bars designate 2 standard deviations for the mean estimate. The blue plot has the highest NRMSE for the central region because there is too much overlap over this region. The orange plot has the highest error for the outermost regions because there is not enough overlap over these regions.}\label{Fig: NRMSE v regions}
\end{figure}

Figure~\ref{Fig: NRMSE v regions} plots the NRMSE for these two geometries as a function of depth of the region of the reconstructed OPL plane, averaged over the 3000 reconstructions.  The narrow 3v,2$^\circ$-geometry reconstructs the OPL contributions from the outermost regions significantly better than the wide 3v,16$^\circ$-geometry. This indicates that the views for the wider geometry do not overlap enough over the outermost regions. Conversely, the narrow geometry has greater error for the central OPL plane, which indicates that the 3 beams for the narrow geometry overlap too much over this central region. Thus, each beam contributes very little unique information for reconstructing this region.

\subsubsection{Zernike expansion error analysis}
\label{sec: zernike1}
\begin{figure}[t]
\centering
    \includegraphics[width=0.9\textwidth]{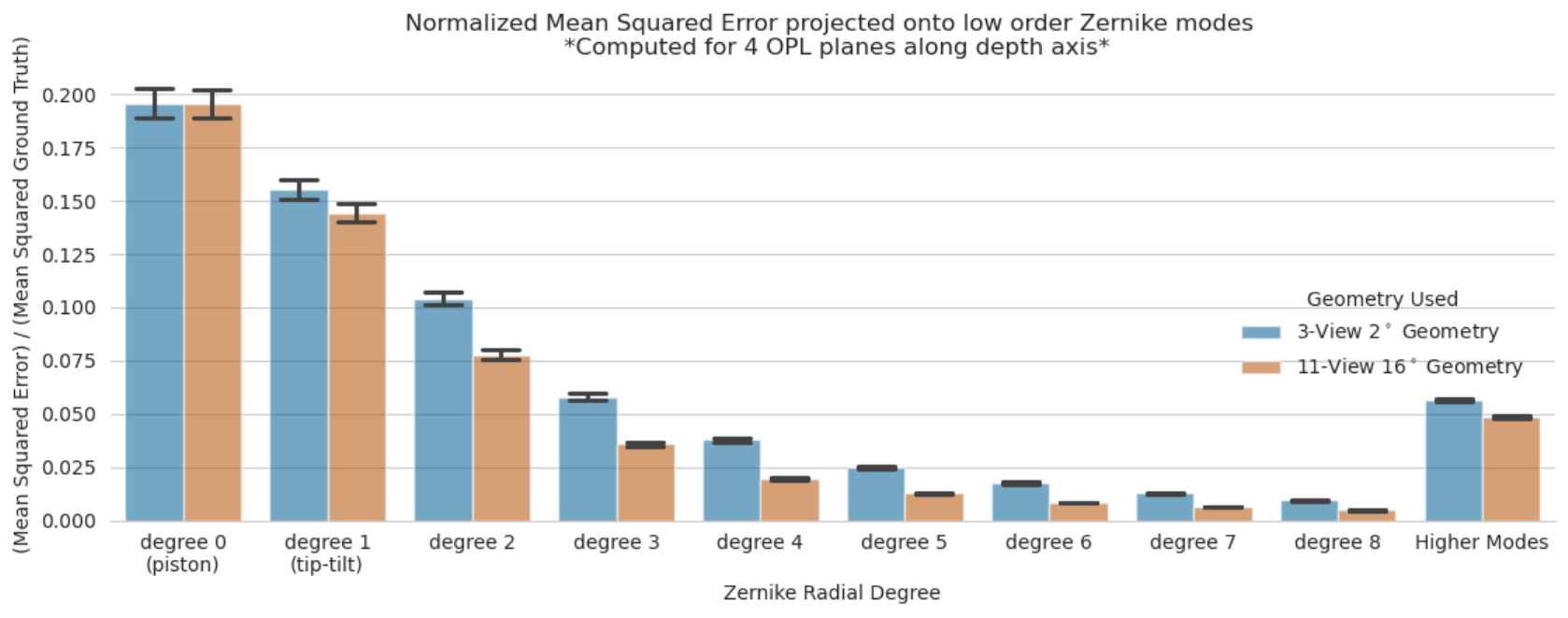}
    \caption{Normalized MSE as a function of Zernike radial degree for reconstructing 4 OPL planes with $\text{OPD}_{\text{TT}}$ measurements, averaged over 1000 samples. The data-rich 11v,16$^\circ$-geometry (brown) shows significantly lower error at high degrees than the data-poor 3v,2$^\circ$-geometry. Both geometries have large errors in degrees 0 and 1, which are not measured in $\text{OPD}_{\text{TT}}$.}\label{Fig: Zernike Distribution}
\end{figure}

Figure~\ref{Fig: Zernike Distribution} shows the normalized mean squared error (MSE) for OPL reconstruction using $\text{OPD}_\text{TT}$ measurements, as a function of Zernike radial degree, for the best (11v,16$^\circ$) and worst (3v,2$^\circ$) geometries of those we investigated (see Fig.~\ref{Fig: Geometries}).  The normalized MSE is $\sum_j \|\hat{\bx}_{d,j}^{err}\|^2 / \sum_j \|\bx_j\|^2$ as described in Sec.~\ref{sec: zernike_expansion}, and we average over 1000 samples. Note that since low order degrees tend to contain more energy in general, we expect to see some energy decay from low to high order degrees.

Figure~\ref{Fig: Zernike Distribution} indicates that lowest order Zernike modes (particularly TTP) contain the largest percentage of the error energy. The TTP modes account for approximately 60\% of the overall MSE for the 11v,16$^\circ$-geometry (brown) and 51\% for the 3v,2$^\circ$-geometry (blue).
The large errors in TTP modes are expected, given that the $\text{OPD}_{\text{TT}}$ measurements do not detect TTP. This conclusion is further validated by the fact that the better geometry only marginally outperforms the worse geometry for the TTP modes. 

\begin{figure}[t]
\centering
    \includegraphics[width=0.98\textwidth]{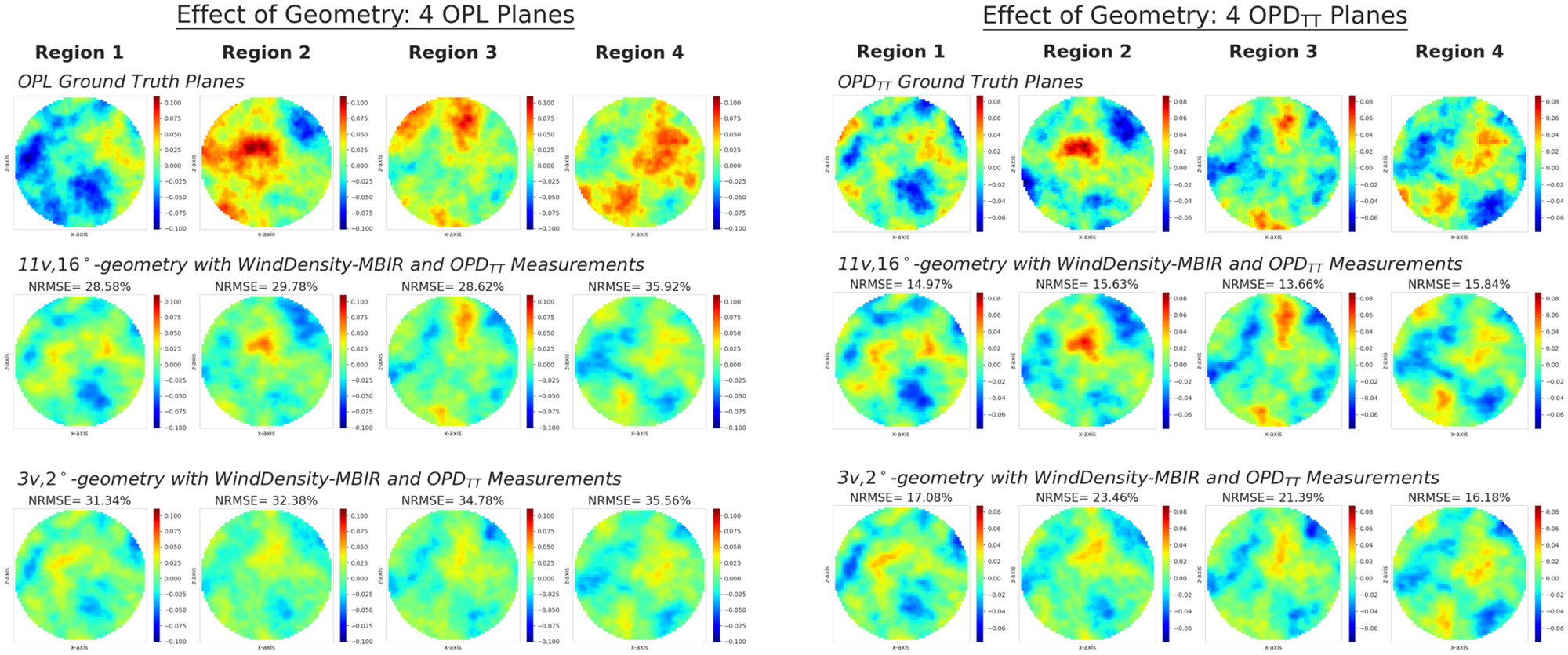}
    \\
    (a) \hspace{7.5cm} (b)
    \caption{Effect of geometric configuration on reconstruction quality. Example reconstructions of 4 planes: (a) reconstructions of 4 OPL planes and (b) reconstructions of 4 $\text{OPD}_{\text{TT}}$ planes, all using WindDensity-MBIR. The top row shows ground truth, the middle row uses the 11v,16$^\circ$-geometry, and bottom row uses the 3v,2$^\circ$-geometry.}\label{Fig: Example Recon Min Max}
\end{figure}

Figure~\ref{Fig: Example Recon Min Max} compares reconstructed planes to ground truth planes. Display (a) shows the OPL contributions from each region, while display (b) shows the $\text{OPD}_{\text{TT}}$ contributions from each region. For both (a) and (b) the top row is the ground truth, the middle row is the reconstruction using the best geometry, and the bottom row is the reconstruction using the worst geometry. We see that the visual match and NRMSE are much better when we reconstruct the $\text{OPD}_{\text{TT}}$ rather than OPL, which contains unmeasured TTP information.

\subsubsection{Depth resolution error analysis}

\begin{figure}[t]
\centering
    \includegraphics[width=0.8\textwidth]{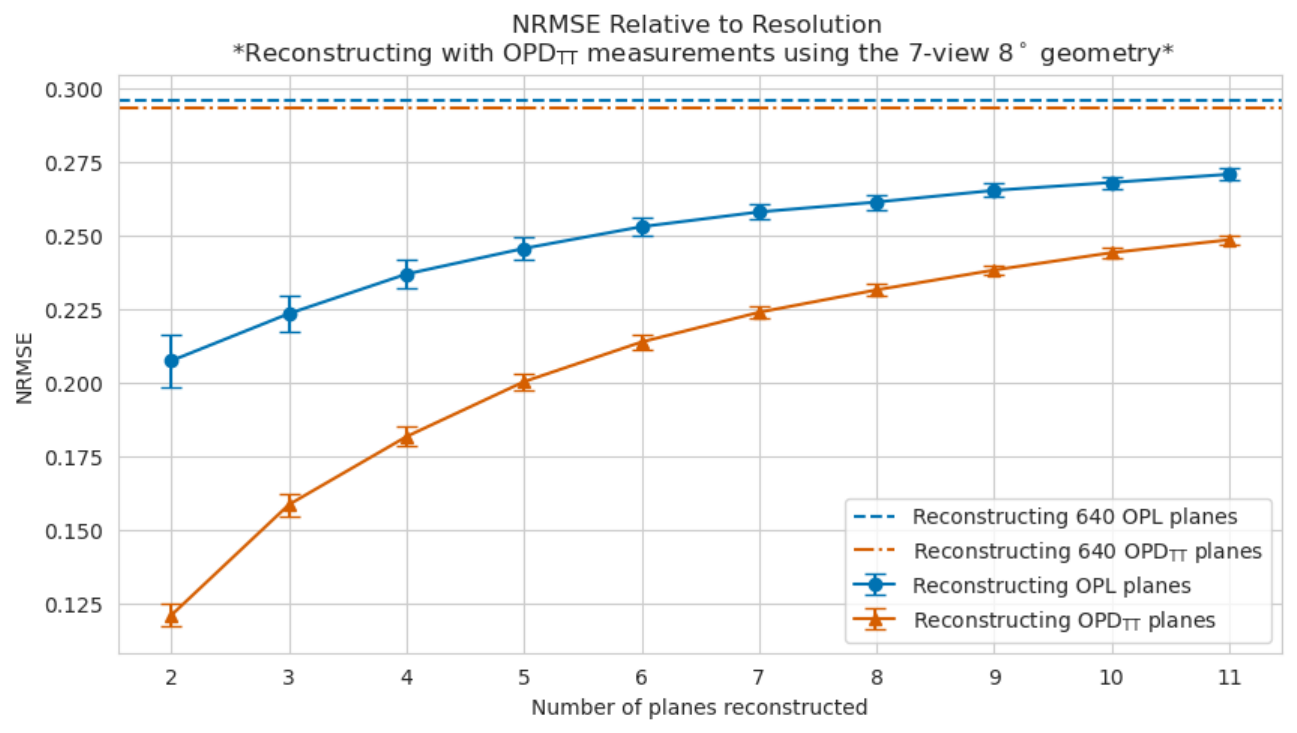}
    \caption{NRMSE (averaged over 100 samples) as a function of number of reconstruction planes using the 7v,8$^\circ$-geometry and $\text{OPD}_{\text{TT}}$ measurements.  The blue curve shows OPL reconstruction, while the orange shows $\text{OPD}_{\text{TT}}$ reconstruction. Due to the missing TTP information, the OPL reconstruction shows consistently higher error, and both curves show a significant increase in error from 2 to 5 planes. }\label{Fig: NRMSE vs Resolution}
\end{figure}

Figure~\ref{Fig: NRMSE vs Resolution} plots the average reconstruction NRMSE relative to the number of OPL or $\text{OPD}_{\text{TT}}$ planes along the depth axis. The plots are the result of performing 100 different reconstructions with the 7v,8$^\circ$-geometry and then computing the NRMSE for various resolutions along the depth axis.  We can see that the reconstruction quality degrades quickly as we increase the resolution from 2 to 5 planes.

 Figure~\ref{Fig: NRMSE vs Resolution} indicates that, to achieve less than 20\% error, we can expect to accurately reconstruct at most 4 $\text{OPD}_{\text{TT}}$ planes. Based on qualitative experience, we have found that when the error exceeds 20\%, the overall visual match starts to degrade significantly.

\subsection{Model Mismatch Effects from Using Tip-tilt Removed Measurements}
\label{sec: OPL Results}
For our secondary analysis, we attempt to quantify the error due to model mismatch from using $\text{OPD}_{\text{TT}}$ measurements instead of OPL. To do this, we perform additional reconstructions with simulated ideal OPL measurements and compare to our earlier reconstructions with $\text{OPD}_{\text{TT}}$ measurements. The main conclusion from our investigation is that the vast majority of the additional error due to model mismatch from using non-ideal $\text{OPD}_{\text{TT}}$ measurements is contained in the low order TTP modes (radial degrees 0 and 1) of the reconstruction, with some small additional error contained in radial degree 2. In particular, for solely recovering $\text{OPD}_{\text{TT}}$ planes along the depth axis, reconstructing with $\text{OPD}_{\text{TT}}$ measurements is only marginally worse than reconstructing with OPL.

\subsubsection{Zernike analysis}
\label{sec: zernike2}
\begin{figure}[t]
\centering
    \includegraphics[width=0.9\textwidth]{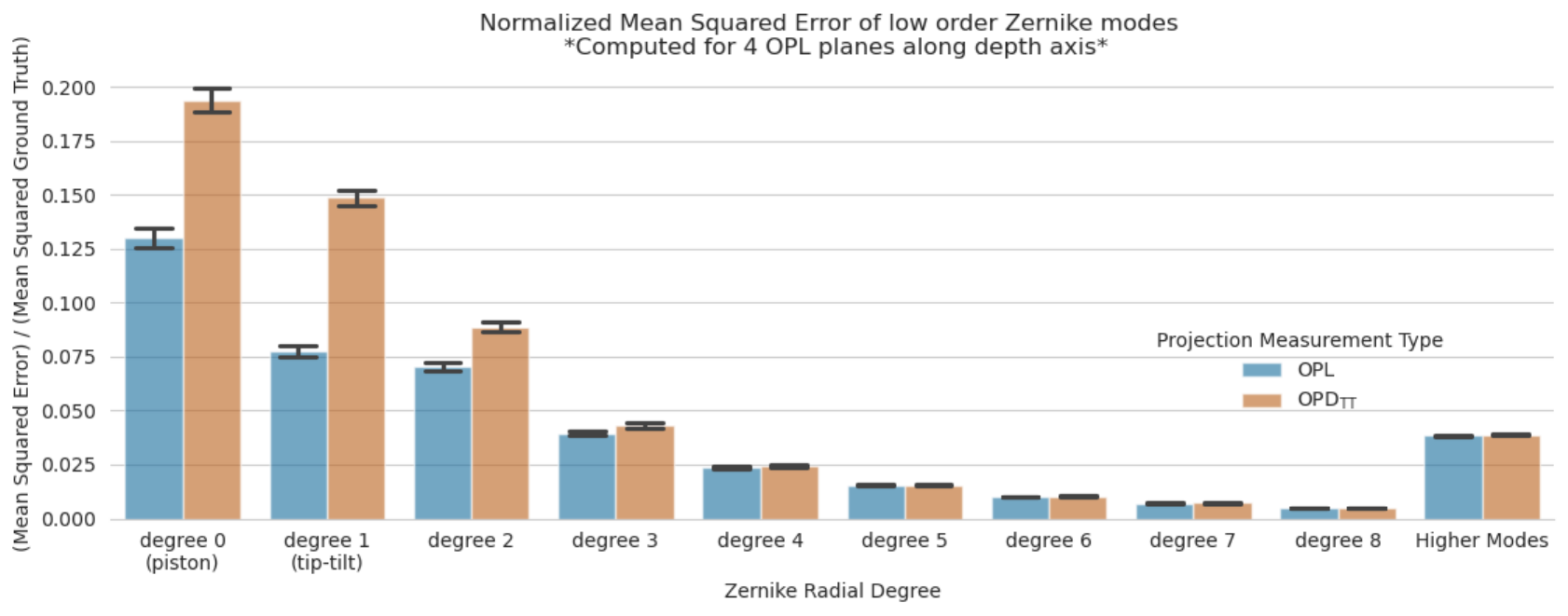}
    \caption{Results analogous to Fig.~\ref{Fig: Zernike Distribution} for reconstructing 4 OPL planes, but comparing the effect of OPL measurements versus $\text{OPD}_{\text{TT}}$ measurements.  The primary difference in error is in the low order degrees 0 and 1, which are not measured in $\text{OPD}_{\text{TT}}$.  Roughly 95\% of the additional error from removing TTP (model-mismatch) is concentrated in degrees 0, 1, 2.
    }\label{Fig: Zernike Distribution OPL and OPD}
\end{figure}

Figure~\ref{Fig: Zernike Distribution OPL and OPD} shows the Zernike modal energy distribution of the normalized mean squared reconstruction error for both reconstructing with OPL measurements and reconstructing with $\text{OPD}_{\text{TT}}$ measurements. For this result, we used the 7v,8$^\circ$-geometry. We can see that Zernike modes of radial degree 0 and 1 (i.e., TTP) contain most of the additional error, which we should expect since the $\text{OPD}_{\text{TT}}$ measurements do not included TTP. To be precise, on average 80\% of the additional mean squared error due to using $\text{OPD}_{\text{TT}}$ measurements instead of OPL is contained in Zernike modes of radial degree 0 and 1. Furthermore, the majority of the higher order effects due to model mismatch are accounted for by radial degree 2, which on average contained 15\% of the difference in MSE between OPL tomography and $\text{OPD}_{\text{TT}}$ tomography.

\begin{figure}[t]
\centering
    \includegraphics[width=0.98\textwidth]{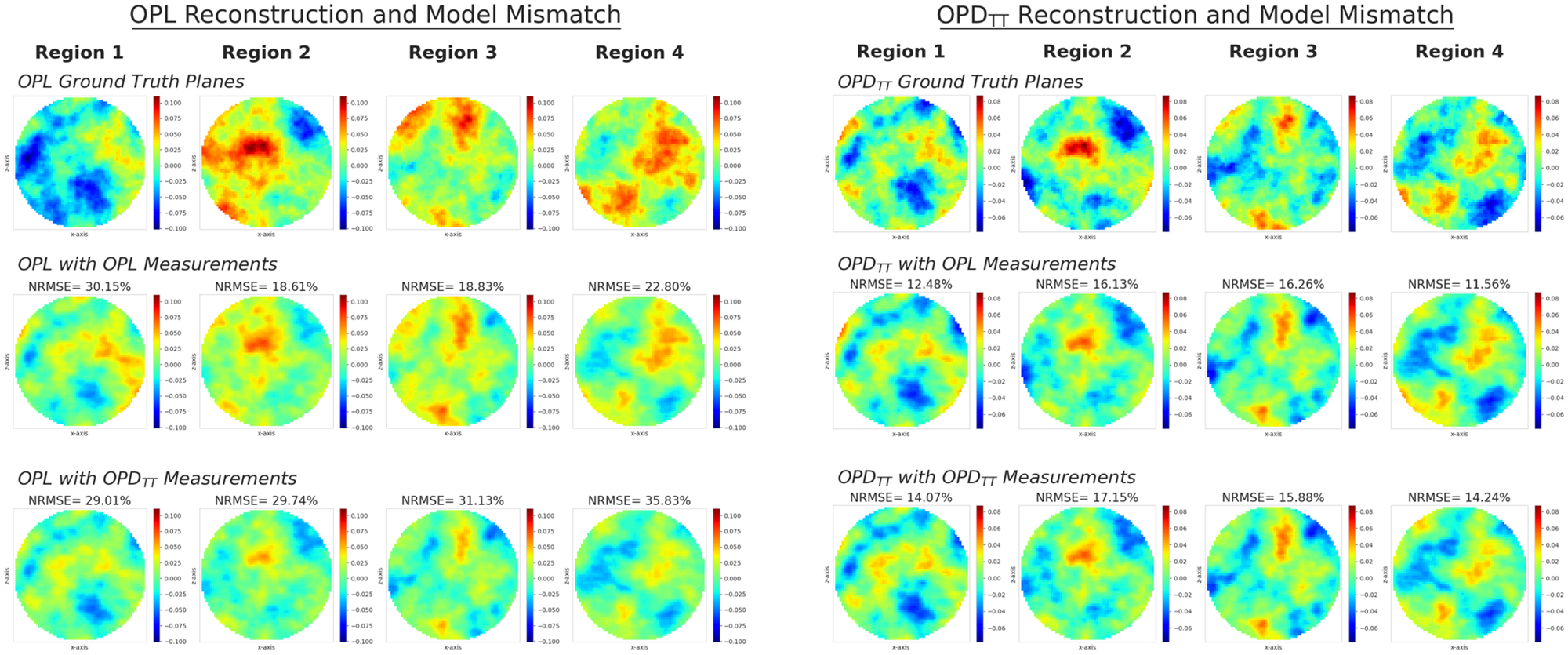}
    \newline
    (a) \hspace{7.5cm} (b)
    \caption{Effect of OPL versus $\text{OPD}_{\text{TT}}$ measurements.  Example reconstructions of 4 planes along the depth axis using the 7v,$8^\circ$ geometry: (a) reconstruction of 4 OPL planes and (b) reconstruction of 4 $\text{OPD}_{\text{TT}}$ planes. The top row shows ground truth, the middle row uses OPL measurements, and the bottom row uses $\text{OPD}_{\text{TT}}$ measurements.  Note the TTP information in OPL measurements significantly improves the OPL reconstructions but only marginally improves the $\text{OPD}_{\text{TT}}$ reconstructions.}
    \label{Fig: Example Recon Comparing OPL to OPD tomography}
\end{figure}

Figure~\ref{Fig: Example Recon Comparing OPL to OPD tomography} compares (a) OPL and (b) $\text{OPD}_{\text{TT}}$ reconstructions using OPL measurements and $\text{OPD}_{\text{TT}}$ measurements.  We see that the OPL reconstructions using $\text{OPD}_{\text{TT}}$ measurements failed to capture certain features that were captured in OPL reconstructions using $\text{OPL}$ measurements. However, the $\text{OPD}_{\text{TT}}$ reconstructions using either $\text{OPL}$ or $\text{OPD}_{\text{TT}}$ look nearly identical and have comparable NRMSE.

\subsubsection{Resolution analysis} 
\begin{figure}[t]
\centering
    \includegraphics[width=0.98\textwidth]{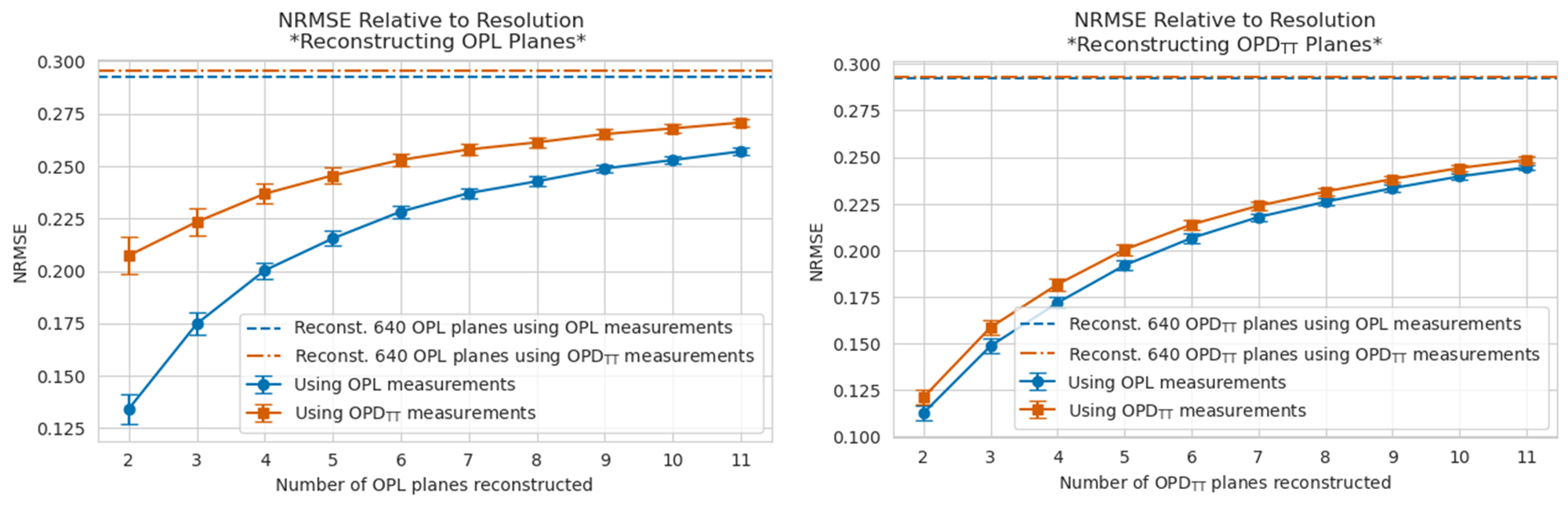}

    (a) \hspace{7.5cm} (b)
    \caption{NRMSE as a function of depth axis resolution: (a) NRMSE for reconstructing OPL planes and (b) NRMSE for reconstructing $\text{OPD}_{\text{TT}}$ planes. For reconstructing $\text{OPD}_{\text{TT}}$ planes, regardless of resolution, using $\text{OPD}_{\text{TT}}$ measurements is only marginally worse than using OPL measurements.}\label{Fig: NRMSEvres_OPLandOPD}
\end{figure}

Figure~\ref{Fig: NRMSEvres_OPLandOPD} plots the NRMSE as a function of depth axis resolution for the case of using OPL measurements (blue) and the case of using $\text{OPD}_{\text{TT}}$ measurements (orange). Plot (a) shows the NRMSE for reconstructing OPL, while plot (b) shows the NRMSE for reconstructing $\text{OPD}_{\text{TT}}$ planes, all using the 7v,8$^\circ$-geometry. The difference between plots (a) and (b) of Fig.~\ref{Fig: NRMSEvres_OPLandOPD} confirms that, regardless of resolution, if we reconstruct $\text{OPD}_{\text{TT}}$ planes, then using OPL measurements is only marginally worse than using $\text{OPD}_{\text{TT}}$ measurements.

\section{Conclusion}
Wind tunnels offer a controlled setting for studying aerodynamic turbulence, but obtaining non-invasive 3D density measurements remains challenging. Most basic techniques, such as optical path difference imaging, provide only 2D data and miss the 3D turbulence structure. Invasive methods like flow seeding are difficult to perform, while CFD models often diverge from experiments. This highlights the need for tomographic approaches that non-invasively measure the 3D density field.

In this paper, we introduced WindDensity-MBIR, an MBIR tomography method for reconstructing the 3D refractive index inside a wind tunnel. Our simulations show that WindDensity-MBIR can reconstruct the non-TTP (higher order) features of 2 to 6 regions along the depth axis with error between 10\% to 25\%, on average, assuming sufficient views and angular extent. Increasing both the number of views and angular extent simultaneously improves results, but increasing angular extent without increasing the number of views can make the reconstruction quality worse. For our secondary analysis we compared reconstructing with OPL views to tip-tilt removed OPD views, and this indicated that the additional error due to model mismatch (from using non-ideal tip-tilt removed OPD) is contained primarily in lower order Zernike modes, while higher-order features remain recoverable. These findings highlight the potential for model-based methods to perform non-invasive density measurements.

\section{Disclosures} 

The authors declare that there are no financial interests, commercial affiliations, or other potential conflicts of interest that could have influenced the objectivity of this research or the writing of this paper.

The views expressed are those of the author and do not necessarily reflect the official policy or position of the Department of the Air Force, the Department of Defense, or the U.S. Government. Approved for public release; distribution is unlimited. Public Affairs release approval \# \PAnumber .

\section* {Code, Data, and Materials Availability} 
The code used to generate the figures and results is available in a Github repository linked in Ref.~\citenum{Repo}. The simulated data used to validate our method is not available due to memory storage constraints.

\section* {Acknowledgments}
C.A.B. was partially supported by the Showalter Trust.  K.J.W., G.T.B., and C.A.B. were partially supported by AFRL/RDKL FA9451-20-2-0008. The authors would like to thank the Showalter family and the United States Air Force for supporting this research.

\bibliography{report}  
\bibliographystyle{spiejour} 

\vspace{2ex}\noindent\textbf{Karl J. Weisenburger} is Mathematics PhD candidate at Purdue University. He received his BS in applied mathematics from Hillsdale College in 2022. His current research interests include computational imaging and aero-optics.

\vspace{1ex}

\listoffigures
\listoftables
\end{spacing}
\end{document}